\documentclass[doublespace,a4paper,12pt]{article}
\usepackage[latin1]{inputenc}
\usepackage{amsmath}
\usepackage{amsfonts}
\usepackage{amssymb}
\usepackage{multirow}\usepackage{rotating}
\usepackage[unicode=true, breaklinks=true, pdfborder={0 0 0}, colorlinks=true, linkcolor=black, citecolor=blue]{hyperref}
\usepackage{lscape}
\usepackage{sidecap}
\usepackage{geometry}

\newcommand{\ud}{\mathrm{d}}
\newcommand{\w}{\tilde}

\author{Joel Berkeley, David Berman}
\title{The Navier-Stokes equation and solution generating symmetries from holography}

\begin{document}

\begin{titlepage}

\vfill
\begin{flushright}
QMUL-PH-12-15
\end{flushright}

\vfill

\begin{center}
   \baselineskip=16pt
   {\Large \bf The Navier-Stokes equation and solution generating symmetries from holography}
   \vskip 2cm
Joel Berkeley\footnote{\tt j.a.fitzhardinge-berkeley@qmul.ac.uk} and David S. Berman\footnote{\tt d.s.berman@qmul.ac.uk},
       \vskip .6cm
             \begin{small}
                          {\it Queen Mary University of London, Centre for Research in String Theory, \\
             School of Physics, Mile End Road, London, E1 4NS, England} \\ 
\vspace{2mm}
\end{small}
\end{center}

\begin{abstract}
The fluid-gravity correspondence provides us with explicit spacetime metrics that are holographically dual to (non-)relativistic nonlinear hydrodynamics. The vacuum Einstein equations, in the presence of a Killing vector, possess solution-generating symmetries known as spacetime Ehlers transformations. These form a subgroup of the larger generalized Ehlers group acting on spacetimes with arbitrary matter content. We apply this generalized Ehlers group, in the presence of Killing isometries, to vacuum metrics with hydrodynamic duals to develop a formalism for solution-generating transformations of Navier-Stokes fluids. Using this we provide examples of a linear energy scaling from RG flow under vanishing vorticity, and a set of $\mathbb{Z}_2$ symmetries for fixed viscosity.
\end{abstract}

\vfill
\setcounter{footnote}{0}
\end{titlepage}

\tableofcontents

\section{Introduction} \label{sec:Fluid-gravity}

In 1974, Damour \cite{Damour}, and later in 1986, Thorne et al. \cite{Membrane Paradigm}, considered an observer outside a black hole, interacting with (perturbing) the event horizon. Surprisingly, they found that the observer will experience the perturbations of the ``stretched" horizon as modes of a viscous fluid possessing electric charge and conductivity. This inspired a host of works over the years but the connection between gravitational physics and fluids became sharper with the advent of the AdS/CFT correspondence when Policastro et. al \cite{Policastro:2001yc} related the shear viscosity of $\mathcal{N}=4$ super Yang-Mills theory to the absorption of energy by a black brane. 

This was the start of using the holographic principle, that is the correspondence between gravitational theories on $(d+1)$-dimensional manifolds and $d$-dimensional quantum field theories, as a tool for calculating hydrodynamic properties.

More recently, there has been a set of works that directly relate solutions of Einstein's equations of a particular type to solutions to the Navier-Stokes equations in one dimension less \cite{Bhattacharyya:2008jc,Bredberg:2010ky,Bredberg:2011jq,Compere:2011dx,Bredberg:2011xw,Cai:2011xv,Bhattacharyya:2008kq,Compere:2012mt,Eling:2012ni}. Later we will review the details of how this correspondence is derived but the essential flavour is as follows. One writes down a very particular ansatz for the metric in $d+1$ dimensions which has undetermined functions, $v^i(x,t),P(x,t)$ and parameter, $\nu$ with the index $i=1,..,d-1$ i.e. over a $(d-1)$ subset of the $d$ spacetime dimensions. Solving the Einstein equations then constrains the functions $v^i(x,t),P(x,t)$ to give a set of second order nonlinear differential equations for $v^i(x,t),P(x,t)$. This set of equations are the Navier-Stokes equations describing a fluid in $d$ dimensions with pressure, $P(x,t)$, fluid velocity field $v^i(x,t)$ and viscosity $\nu$. Thus particular solutions to the Navier-Stokes equations provide particular solutions to the Einstein equations.

Other recent work on the fluid/gravity correspondence may be found in \cite{Padmanabhan:2010rp, Matsuo:2012pi, Chapman:2012my, Huang:2011kj, Rodrigues:2011gg, Nakayama:2011bu, Huang:2011he, Brattan:2011my, Kuperstein:2011fn}.

It has been known since Buchdahl \cite{Buchdahl:1959nk} that for manifolds with isometries Einstein's equations have solution generating symmetries. That is there are ``hidden" symmetries of the equations that map from one solution to the other. In fact there is a vast set of these as described by Ehler \cite{Ehlers:1957zz} and Geroch \cite{Geroch:1970nt}. The question we wish to pose in this paper is whether the solution generating symmetries of Einstein's equations can lead to solution generating symmetries in the Navier-Stokes equations? 

The procedure to determine this will be as follows:
\begin{itemize}

\item{impose a Killing symmetry in a spacetime that admits the metric ansatz corresponding to the Navier-Stokes equations;}

\item{carry out generalised Ehler's transformations that preserve the ansatz;}

\item{determine the induced transformation of the Navier-Stokes data i.e. $v^i,P,\nu$.}

\end{itemize}

The procedure could immediately fail if there were no generalized Ehler's transformations that preserve the metric ansatz required for the fluid/gravity correspondence. We will find that there are a finite set and we will be able to explore the transformations on the Navier-Stokes fields for different choices of Killing directions in spacetime. Along the way we will show that they are not part of the usual spacetime Ehler's transformations and yet they do produce solution generating transformations for the Navier-Stokes fields.

The paper will try to be as self contained as possible and so we begin with a review of the necessary ideas in fluids; the Navier-Stokes equation; the fluid gravity correspondence; and solution generating symmetries in general relativity. We will then carry out the procedure described above for spatial, timelike and null Killing vectors to see what solution generating symmetries they correspond to in the Navier-Stokes equation. We end with some comments and ideas for future work. A reader familiar with the formalism of hydrodynamics and the Navier-Stokes equation may wish to skip directly to section 3 where we carry out the solution generating transformation in the gravity dual to see the resulting induced transformations on the solutions of the Navier-Stokes equation.

\section{Hydrodynamics}

The study of hydrodynamics is fundamental to vast areas of physics and engineering, owing to its origin as the long-wavelength limit of any interacting field theory at finite temperature. Such a limit needs a consistent definition. Consider a quantum field theory where quanta interact with a characteristic length scale $\ell_\text{corr}$, the correlation length. The long-wavelength limit simply requires that fluctuations of the thermodynamic quantities of the system vary with a length scale $L$ much greater than $\ell_\text{corr}$, parameterized by the dimensionless Knudsen number \begin{equation}
K_n=\frac{\ell_\text{corr}}{L}. \label{Knudsen}
\end{equation}
For a fluid description to be useful in non-equilibrium states, we naturally require that $L$ remain small compared to the size of the system. This is usually satisfied trivially by considering systems of infinite size.

The long-wavelength limit allows the definition of a particle as an element of the macroscopic fluid, infinitesimal with respect to the size of the system, yet containing a sufficiently large number of microscopic quanta. One mole contains an Avogadro's number of molecules, for example. Each particle defines a local patch of the fluid in thermal equilibrium, that is, thermodynamic quantities do not vary within the particle. Away from global equilibrium quantities vary between particles as function of time $\tau$ and spatial coordinates $\vec x$, combined as $x^a=(\tau,\vec x)$. The evolution of particles in the fluid is parameterized by a relativistic velocity $u^b(x^a)$, which refers to the velocity of the fluid at $x^a$. It is well known \cite{LL} that the thermodynamic quantities, such as the temperature $T(x^a)$ and the density $\rho(x^a)$, are determined by the value of any two of them, along with the equation of state. The evolution of the system is then specified by the equations of hydrodynamics in terms of a set of transport coefficients, whose values depend on the fluid in question.

Fluid flow is in general relativistic in that the systems it describes are constrained by local Lorentz invariance, and velocities may take any physical values below the speed of light. Applications at relativistic velocities are multitudinous: the dust clouds in galaxy and star formation; the flow of plasmas and gases in stars supporting fusion; the superfluid cores of neutron stars; the horizons of black holes are all described by hydrodynamics. Modelling black holes (and black branes in M/string theory) with hydrodynamics has now developed into a fundamental correspondence of central importance to our present study, as discussed in \S\ref{sec:Fluid-gravity}. Quark-gluon plasmas behave as nearly ideal fluids and are expected to have formed after the inflationary epoch of the big bang, are reproduced in collisions at the RHIC and LHC. Non-relativistic fluids are equally ubiquitous, somewhat more familiar, and constitute an endless list of phenomena from the atmosphere to the oceans.

\subsection{The fluid equations} \label{sec:Relativistic fluids}

We begin with a discussion, adapted from \cite{Rangamani:2009xk}, of the relativistic fluid described by the stress energy tensor $T^{ab}$ and a set of conserved currents $J^a_I$ where $I$ indexes the corresponding conserved charge. The dynamical equations of the $d$-spacetime dimensional fluid are
\begin{equation}
\nabla_a T^{ab}=0	\qquad	\nabla_a J^a_I=0 \, . \label{relativistic NS}
\end{equation}
For an ideal fluid, with no dissipation, the energy-momentum tensor and currents may be expressed in a local rest frame in the form
\begin{subequations}\label{TJideal}\begin{align}
T^{ab} &=\rho u^au^b+p(g^{ab}+u^au^b) \label{Tideal}\\
J_I^a &=q_I u^a \label{Jideal}
\end{align}\end{subequations}
where $p$ is the pressure, $q_I$ are the conserved charges and $g_{ab}$ is the metric of the space on which the fluid propagates. The velocity is normalised to $u^au_a=-1$. The entropy current is given by \eqref{Jideal} with the charge $q$ being given by the local entropy density. The conservation of the entropy current illustrates the non-dissipative nature intrinsic to zero entropy production.

In a dissipative fluid, there are corrections to \eqref{TJideal}. We must first take into account the interrelation between mass and energy to define the velocity field more rigorously. This is achieved by using the Landau gauge, which requires that the velocity be an eigenvector of the stress-energy tensor with eigenvalue the local energy density of the fluid (this is satisfied by the velocity normalisation for the ideal fluid). If the stress energy tensor gains a dissipative term $\Pi^{ab}$, and the current a term $\Upsilon_I^a$, this reads \[\Pi^{ab}u_a=0\qquad \Upsilon_I^au_a=0.\]
Dissipative corrections to the stress tensor are constructed in a derivative expansion of the velocity field and thermodynamic variables, where derivatives implicitly scale with the infinitesimal Knudsen number \eqref{Knudsen}. Recalling that the equations of motion for the ideal fluid are composed of relations between these gradients, we may express $\Pi^{ab}$ purely in terms of the derivative of the velocity (when charges are present this is only true to to first order). This can be iterated to all orders in the expansion. Now, the derivative of the velocity may be decomposed using the acceleration $A^a$, divergence $\theta$, a symmetric traceless shear $\sigma^{ab}$, and the antisymmetric vorticity $\omega^{ab}$ into the form \[\nabla^b u^a=-A^a u^b+\sigma^{ab}+\omega^{ab}+\frac{1}{d-1}\theta P^{ab},\] where
\begin{equation*}\begin{split}
\theta &=\nabla_a u^a\\
A^a &=u^b\nabla_b u^a\\
\sigma^{ab} &=P^{ac}P^{bd}\nabla_{(c}u_{d)}-\frac{1}{d-1}\theta P^{ab}\\
\omega^{ab} &=P^{bc}P^{ad}\nabla_{[c}u_{d]}.
\end{split}\end{equation*}
and $P^{ab}=g^{ab}+u^a u^b$ is a projection operator onto spatial directions.
In the Landau frame, only the divergence and shear can contribute to first-order stress-energy tensor. A similar analysis for the charge current retains the acceleration, and if one includes the parity-violating pseudo-vector contribution \[\ell^a ={\epsilon_{bcd}}^a u^b\nabla^c u^d,\] the leading order dissipative equations of motion for a relativistic fluid are \eqref{relativistic NS} with
\begin{subequations}\label{TJdiss}\begin{align}
T^{ab} &=\rho u^au^b+pP^{ab}-2\eta\sigma^{ab}-\zeta\theta P^{ab}	\label{Tdiss}\\
J_I^a &=q_I u^a	-\chi_{IJ}P^{ab}\nabla_b q_J-\Theta_I\ell^a-\gamma_I P^{ab}\nabla_bT,	\label{Jdiss}
\end{align}\end{subequations}
where $\eta$ and $\zeta$ are the shear\footnote{We use the traditional notation of \cite{Kovtun:2004de, Bhattacharyya:2008kq, LL} rather than that of \cite{Bredberg:2011jq, Compere:2011dx}. We introduce the kinematical viscosity $\nu$ in \S\ref{sec:non-rel NS}.} and bulk viscosities respectively, $\chi_{IJ}$ is the matrix of charge diffusion coefficients, $\gamma_I$ indicates the contribution of the temperature gradients and $\Theta_I$ the pseudo-vector transport coefficients. The transport coefficients have been calculated in the weakly coupled QFT in perturbation theory, whereas in the strongly coupled theory, a dual holographic description may be employed, see e.g. \cite{Baier:2007ix}.

\subsubsection{The incompressible Navier-Stokes equations}\label{sec:non-rel NS}

In the non-relativistic limit defined by long distances, long times and low velocity and pressure amplitudes (see e.g \cite{Bhattacharyya:2008kq}), the fluid equations \eqref{relativistic NS} with \eqref{TJdiss} become the incompressible non-relativistic Navier-Stokes equations. In flat space and in the presence of an external electromagnetic field $a_i$, these are
\begin{subequations}\label{E}\begin{gather}
\partial_\tau v_i-\nu\partial^2 v_i+\partial_i P+v^j \partial_j v_i=-\partial_\tau a_i-v^j f_{ji}, \label{Ei}\\
\partial_iv^i=0, \label{E0}
\end{gather}\end{subequations}
where $f_{ij}=\partial_i a_j-\partial_j a_i$ is the field strength of $a_i$. Ideal fluids are described by Euler's equations, where the \textit{kinematical} viscosity $\nu$ (related to the shear viscosity) vanishes. We will mostly be concerned with fluid flow in the absence of external forces, where $a_i$ is zero.

\subsection{The Navier-Stokes fluid on a Rindler boundary} \label{sec:Rindler fluid}

A metric dual to the non-relativistic incompressible Navier-Stokes equations was first developed in \cite{Bredberg:2011jq} on the Rindler wedge, up to third order in the non-relativistic, small amplitude expansion detailed later in this section. An algorithm for generalising this metric to all orders was subsequently developed in \cite{Compere:2011dx}, though terms calculated beyond third order are not universal. They receive corrections from quadratic curvature in Gauss-Bonnet gravity \cite{Cai:2011xv}. We summarise the construction in \cite{Compere:2011dx} here.

\begin{SCfigure}\centering
\caption{The past $\mathcal{H}^-$ and future $\mathcal{H}^+$ horizons define the boundary of the Rindler wedge. Grey lines demonstrate lines of constant $r$ (curved) and $\tau$ (straight). Long-wavelength perturbations of the hypersurface $\Sigma_c$ are described by the equations of hydrodynamics.\newline\newline\newline\newline\newline}
\includegraphics[scale=0.3]{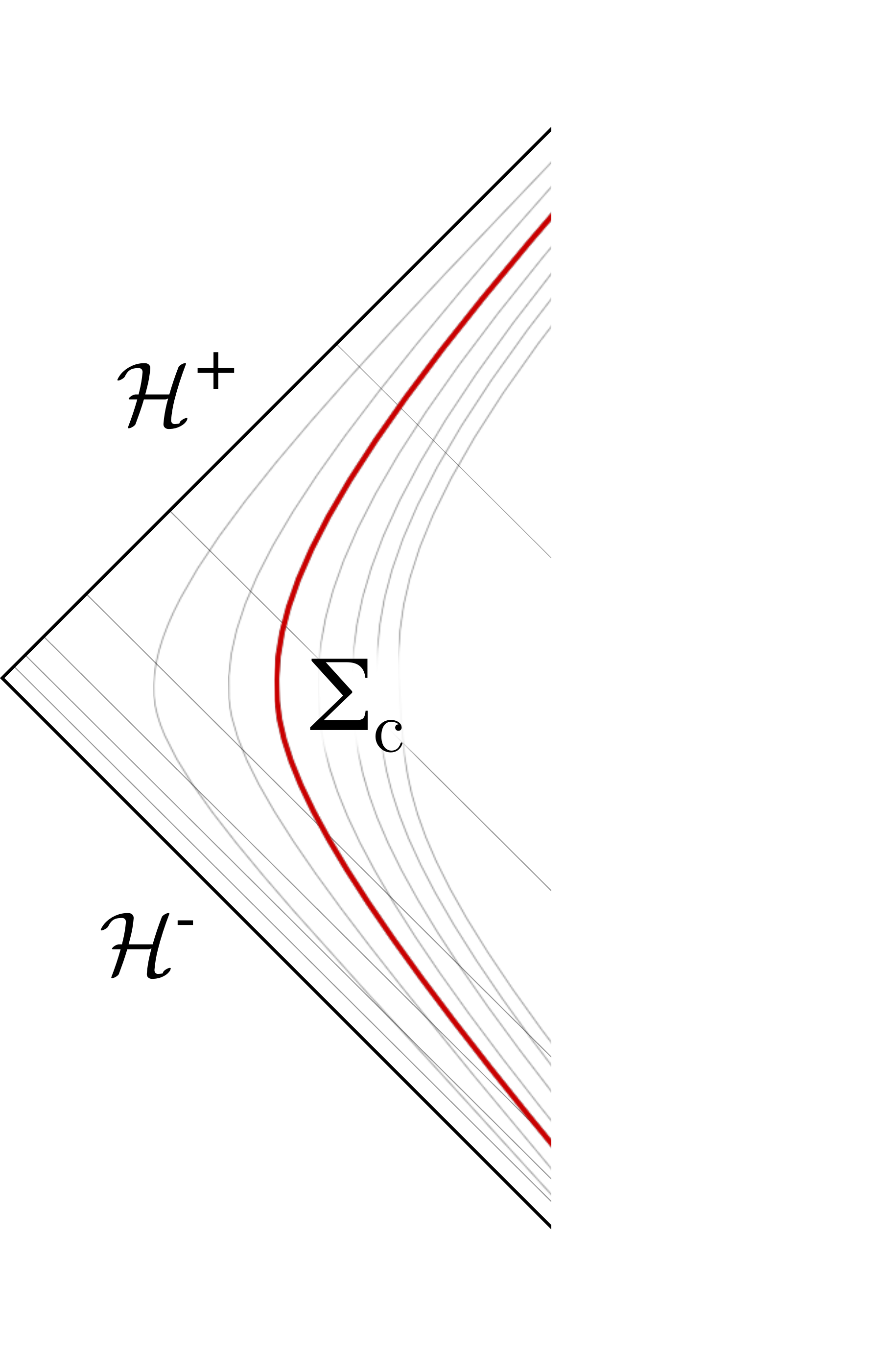}\label{fig:Rindler space}
\end{SCfigure}

Consider the surface $\Sigma_c$ with induced metric
\begin{equation}
\gamma_{ab}\ud x^a\ud x^b=-r_c\ud\tau^2+\ud x_i \ud x^i \label{flat metric}
\end{equation}
where the parameter $\sqrt r_c$ is an arbitrary constant. One metric embedding this surface is
\begin{equation}
\ud s^2=-r\ud\tau^2 +2 \ud\tau\ud r + \ud x_i \ud x^i,
\end{equation}
which describes flat space (fig. \ref{fig:Rindler space})
in ingoing Rindler coordinates $x^\mu=(\tau,x^i,r)$, defined in terms of the Cartesian chart $(t,x^i,z)$ by
\begin{equation}
z^2-t^2=4r \qquad z+t=e^{\tau/2}. \label{rindler coordinates}
\end{equation}
The hypersurface $\Sigma_c$ is defined by  $r=r_c$ where $r$ is the coordinate into the bulk. Allowing for a family of equilibrium configurations, consider diffeomorphisms satisfying the three conditions
\begin{itemize}
\item[i)] The induced metric on the hypersurface $\Sigma_c$ takes the form \eqref{flat metric}
\item[ii)] The stress tensor on $\Sigma_c$ describes a perfect fluid
\item[iii)] Diffeomorphisms return metrics stationary and homogeneous in $(\tau,x^i)$.
\end{itemize}
The allowed set is reduced to the following boost, shift and rescaling of $x^\mu$. First, a constant boost $\beta_i$,
\begin{equation}
\sqrt{r_c}\tau\rightarrow\gamma(\sqrt{r_c}\tau-\beta_i x^i), \qquad x^i\rightarrow x^i-\gamma\beta^i\sqrt{r_c}\tau+(\gamma-1)\frac{\beta^i\beta_j}{\beta^2}x^j,\label{rindlerdiffeo1}
\end{equation}
where $\gamma=(1-\beta^2)^{-1/2}$ and $\beta_i\equiv {r_c}^{-1/2}v_i$. Second, a shift in $r$ and a rescaling of $\tau$,
\begin{equation}
r\rightarrow r-r_h, \qquad \tau\rightarrow(1-r_h/r_c)^{-1/2}\tau.\label{rindlerdiffeo2}
\end{equation}
These yield the flat space metric in rather complicated coordinates,
\begin{equation}\begin{split}
\ud s^2 =& \frac{\ud \tau^2}{1-v^2/r_c}\left(v^2-\frac{r-r_h}{1-r_h/r_c}\right)-\frac{2\gamma}{\sqrt{1-r_h/r_c}}\ud\tau\ud r - \frac{2\gamma v_i}{r_c\sqrt{1-r_h/r_c}}\ud x^i\ud r\\
 &+ \frac{2 v_i}{1-v^2/r_c}\left(\frac{r-r_c}{r_c-r_h}\right)\ud x^i \ud\tau + \left(\delta_{ij}-\frac{v_i v_j}{r_c^2(1-v^2/r_c)}\left(\frac{r-r_c}{1-r_h/r_c}\right)\right)\ud x^i\ud x^j.  \label{metricby}
\end{split}\end{equation}

The Brown-York stress tensor on $\Sigma_c$ (in units where $16\pi G=1$) is given by
\begin{equation}
T_{ab}=2(K\gamma_{ab}-K_{ab}),
\end{equation}
where \[K_{ab}=\dfrac{1}{2}(\mathcal{L}_n \gamma)_{ab},\qquad K=K^a{}_a,\] are the extrinsic curvature and its mean, and $n^\mu$ is the spacelike unit normal to the hypersurface.

By imposing that the Brown-York stress tensor on $\Sigma_c$ gives that of the stress-energy tensor of a fluid we can identify the parameters of the metric (\ref{metricby})  with the density, $\rho$, pressure, $P$ and four-velocity $u^a$ of a fluid, as follows:
\begin{equation}
\rho=0, \qquad p=\frac{1}{\sqrt{r_c-r_h}}, \qquad u^a=\frac{1}{\sqrt{r_c-v^2}}(1,v_i).
\end{equation}
 The Hamiltonian constraint \[R_{\mu\nu}n^\mu n^\nu=0\] on $\Sigma_c$ yields a constraint on the Brown-York stress tensor \[dT_{ab}T^{ab}=(T^a_a)^2.\] When this constraint is applied to the equilibrium configurations described above, one finds the equation of state is $\rho=0$ (as above), or $\rho=-2d(d-1)p$ which occurs for a fluid on the Taub geometry \cite{Eling:2012ni}. 

Promoting $v_i$ and $p$ to slowly varying functions of the coordinates $x^a$, and regarding $v_i(\tau,x_j)$ and $p=r_c^{-1/2}+r_c^{-3/2}P(\tau,x_i)$ as small perturbations, which scale as
\begin{equation}
v_i\sim\epsilon, \qquad P\sim\epsilon^2,
\end{equation}
about equilibrium yields the metric
\begin{equation}\begin{split}
\ud s^2 =& -r\ud\tau^2 + 2\ud\tau \ud r + \ud x^i \ud x_i \\
&- 2\left(1-\frac{r}{r_c}\right)v_i \ud x^i \ud\tau - 2\frac{v_i}{r_c}\ud x^i \ud r \\
& +\left(1-\frac{r}{r_c}\right)\left[(v^2+2P)\ud\tau^2 + \frac{v_i v_j}{r_c}\ud x^i \ud x^j\right] + \left(\frac{v^2+2P}{r_c}\right)\ud\tau \ud r +\mathcal{O}(\epsilon^3)
\end{split}\end{equation}
which satisfies the Einstein's equations to $\mathcal{O}(\epsilon^3)$ if $v_i$ satisfies incompressibility, $\partial_i v^i=\mathcal{O}(\epsilon^3)$. Corrections appear in powers of $\epsilon^2$, so this is the complete metric to second order. 

The metric may now be built up order by order in the hydrodynamic scaling. Assume one has the metric at order $\epsilon^{n-1}$, where the first non-vanishing component $\hat R_{\mu\nu}^{(n)}$ of the Ricci tensor appears at order $n$. By adding a correction term $g_{\mu\nu}^{(n)}$ to the metric at order $n$, resulting in a shift in the Ricci tensor $\delta R_{\mu\nu}^{(n)}$, and requiring
\begin{equation}
\hat R_{\mu\nu}^{(n)}+\delta R_{\mu\nu}^{(n)}=0, \label{diff in corrections}
\end{equation}
the vanishing of the Ricci tensor is guaranteed to order $n$. Recalling that, in the hydrodynamic scaling, derivatives scale thus,
\begin{equation}
\partial_r\sim\epsilon^0, \qquad\partial_i\sim\epsilon^1, \qquad \partial_\tau\sim\epsilon^2,
\end{equation}
one sees that corrections $\delta R_{\mu\nu}^{(n)}$ at order $n$ will appear only as $r$ derivatives of $g_{\mu\nu}^{(n)}$. It is shown in \cite{Compere:2011dx} that, using the Bianchi identity and the Gauss-Codacci relations, integrability of the set of differential equations \eqref{diff in corrections} defining $\delta R_{\mu\nu}^{(n)}$ in terms of $g_{\mu\nu}^{(n)}$ is given by imposing the momentum constraint, equivalent to the conservation of the stress tensor on $\Sigma_c$,
\begin{equation}
R_{a\mu}n^\mu=\nabla_a T^{ab}|^{(n)}_{\Sigma_c}=0,
\end{equation}
which is precisely the fluid equations of motion, to order $n$.

The perturbation scheme contains several degrees of freedom. The gauge freedom of the infinitesimal perturbations \[g_{\mu\nu}^{(n)}\rightarrow g_{\mu\nu}^{(n)}+\partial_\mu \varphi_\nu^{(n)}+\partial_\nu \varphi_\mu^{(n)}\] for some arbitrary vector $\varphi^{\mu(n)}(\tau,\vec x,r)$ at order $\epsilon^n$, which may be fixed by demanding that $g_{r\mu}$ is that of the seed metric to all orders in $\epsilon$. The $x^a$-dependent functions of integration from equation \eqref{diff in corrections} may be fixed by imposing the boundary form \eqref{flat metric} of the metric on $\Sigma_c$, and also requiring regularity of the metric at $r=0$, which in this construction translates to the absence of logarithmic terms in $r$. Corrections to the bulk metric under these conditions then become
\begin{equation}\begin{split}
g_{r\mu}^{(n)}=& 0\\
g_{\tau\tau}^{(n)}=& (1-r/r_c)F_\tau^{(n)}(\tau,\vec x)+\int_r^{r_c}\,\ud r' \int_{r'}^{r_c}\,\ud r'' (\hat R_{ii}^{(n)}-r\hat R_{rr}^{(n)}-2\hat R_{r\tau}^{(n)})\\
g_{\tau i}^{(n)}=& (1-r/r_c)F_i^{(n)}(\tau,\vec x)-2\int_r^{r_c}\,\ud r'\int_{r'}^{r_c}\,\ud r'' \hat R_{ri}^{(n)}\\
g_{ij}^{(n)}=& -2\int_r^{r_c}\,\ud r' \frac{1}{r'}\int_0^{r'} \,\ud r''\hat R_{ij}^{(n)},
\end{split}\end{equation}
where the $F_a^{(n)}(\tau,\vec x)$ comprise of the remaining  integration functions, and the final degree of freedom; field redefinitions of $\delta v_i^{(n)}$ and $\delta P^{(n)}$ at order $\epsilon^n$. $F_i^{(n)}(\tau,\vec x)$ is related to redefinitions of the fluid velocity and is fixed by the isotropic gauge condition $P_a^bT_{bc}u^c=0$. $F_\tau^{(n)}(\tau,\vec x)$ is related to redefinitions of the pressure and is fixed by defining the isotropic part of $T_{ij}$ to be
\begin{equation}
T_{ij}^\text{isotropic}=\left(\frac{1}{\sqrt{r_c}}+\frac{P}{r_c^{3/2}}\right) \delta_{ij}
\end{equation}
to all orders.

Applying the perturbation scheme to the seed metric yields to third order,
\begin{equation}
\begin{split}
\ud s^2 =&  -r\ud\tau^2 +2\ud\tau \ud r + \ud x^i \ud x_i - 2\left(1-\frac{r}{r_c}\right)v_i \ud x^i \ud\tau - 2\frac{v_i}{r_c}\ud x^i \ud r \\
& +\left(1-\frac{r}{r_c}\right)\left[(v^2+2P)\ud\tau^2 + \frac{v_i v_j}{r_c}\ud x^i \ud x^j\right] + \left(\frac{v^2+2P}{r_c}\right)\ud\tau \ud r \\
& -\left[\frac{(r^2-r_c^2)}{r_c}\partial^2 v_i + \left(1-\frac{r}{r_c}\right)\left(\frac{v^2+2P}{r_c}\right)v_i\right]\ud x^i \ud\tau + \mathcal{O}(\epsilon^4) \label{g}
\end{split}
\end{equation}
which satisfies the vacuum Einstein equations if
\begin{equation}
r_c^{3/2}\nabla^a T_{ai}|_{\Sigma_c}=\partial_\tau v_i-r_c\partial^2 v_i+\partial_i P+v^j\partial_j v_i=\mathcal{O}(\epsilon^5)
\end{equation}
which are the Navier-Stokes equations with kinematical viscosity
\begin{equation}
\nu=r_c.
\end{equation}

The corresponding corrections to the Navier-Stokes equations follow from conservation of the stress tensor on $\Sigma_c$. Vector and scalar quantities are odd and even orders respectively in the scaling $\epsilon$. Accordingly, corrections to the scalar incompressibility equation appear at even orders, and to the vector Navier-Stokes equations at odd orders.

\section{Duality in the context of holography}

The defining equations in general relativity are the Einstein field equations, and in the non-relativistic limit of hydrodynamics, the Navier-Stokes equations \eqref{E}. Each is a set of non-linear partial differential equations whose solutions exhibit fantastically varied phenomenology. When approaching any complex physical system with a view to finding solutions, it is often advantageous to consider the symmetries, intensively studied in both of these systems since their conceptions. Beyond diffeomorphisms, the search in gravity has in general been somewhat limited \cite{Torre:1993jm, Hoenselaers}, however in the presence of a spacetime isometry, the symmetry group becomes remarkably large \cite{D Maison}, particularly for vacuum spacetimes. For symmetries of the Navier-Stokes equations see \cite{Symmetries of NS}, and with regards to the conformal group \cite{Bhattacharyya:2008kq, Horvathy:2009kz}. In the light of the fluid/gravity correspondence, one may ask whether the symmetries of these systems are linked. In \cite{Stephani88, Garfinkle:1996ur, Boonserm:2006vr}, they apply known symmetries of the Einstein equations to spacetimes with perfect fluid sources, constructing new spacetimes with the same equation of state, though not within a holographic framework. By drawing on the tools provided by these works and those in holography, we hope to develop a more general approach to the problem.

The bulk provides an additional valuable degree of freedom, where the boundary sets the scene for the fluid evolution on the induced geometry. Moreover, we are now free to exploit the symmetries of the more extensive yet simpler vacuum geometries. It is these symmetries which we intend to holographically project to the fluid. In particular, we are interested in transformations between solutions to the Navier-Stokes equations arising from transformations between solutions to the vacuum Einstein equations: transformed metrics yield transformed fluid configurations.

In this section, we discuss the work leading up to the spacetime Ehler's symmetry group of the vacuum Einstein equations with zero cosmological constant, itself contained within the generalized Ehlers group. We continue in \S\ref{sec:SGT in NS} to apply the latter, in the presence of a Killing isometry, to fluids on the boundary of the Rindler space, thus deriving solution generating transformations of the fluid velocity, pressure and viscosity (the latter defining the RG flow). We offer in \S\ref{sec:RG duality} a selection of example transformations including RG flow for zero vorticity fluids (where one may relax this constraint), and $\mathbb{Z}_2$ transformations for fixed viscosity which we show in \S\ref{group for fixed rc} in fact lie outside the spacetime Ehler's group.

\subsection{Symmetry groups of the Einstein equations} \label{sec:Ehlers}

Understanding the properties of the Einstein field equations has long been a subject of great theoretical interest, a sensible starting point being the inherent symmetries involved. To this end, Buchdahl \cite{Buchdahl:1959nk} derived a form of duality in vacuum spacetime metrics, where an $n$-dimensional vacuum metric static with respect to a coordinate $x^s$: \[g_{\mu\nu,s}=0,\qquad g_{a s}=0\qquad\mu,\nu\in\{0,\ldots,n\}\] generates a dual vacuum metric \[h_{\mu\nu}=((g_{ss})^{2/(n-3)}g_{ab},(g_{ss})^{-1}),\qquad a,b\in\{\mu\neq s\}.\] It is this solution-generating property of spacetime isometries we wish to apply to solutions of the Einstein equations and holographically map to hydrodynamics. We have, however, a considerably larger symmetry group at our disposal. The authors of \cite{Mars:2001gd, Ehlers:1957zz, Geroch:1970nt} develop the concept culminating in the generalized Ehlers symmetry group of the Einstein equations also for non-vacuum spacetimes. An extension exists \cite{Woolley, Harrison} to dualities between vacuum spacetimes and those with electromagnetic backgrounds described by the Einstein-Maxwell equations, of relevance for magnetohydrodynamics.

\subsection{The Ehlers group}\label{sec:ehlers}

\subsubsection*{The generalized Ehlers group}

Define a vector field $\xi=\xi^\mu\partial_\mu$ and one-form $W=W_\mu\ud x^\mu$ on a manifold with metric $g=(g_{\mu\nu})$. The \textit{generalized} Ehlers group is defined in \cite{Mars:2001gd} by the transformation
\begin{equation}
g_{\mu\nu}\rightarrow h_{\mu \nu} (\xi,W,g)=\Omega^2 g_{\mu\nu}-2\xi_{(\mu}W_{\nu)}-\frac{\lambda}{\Omega^2}W_\mu W_\nu, \label{ehler}
\end{equation}
where $\Omega^2\equiv\xi^\alpha W_\alpha+1\geq 1$, and the inequality holds over the whole geometry. This group does not send vacuum metrics to vacuum metrics in general, but such transformations may be found in the spacetime Ehlers subgroup.

\subsubsection*{The spacetime Ehlers group}

Let us restrict $g$ to be some $(3+1)$-dimensional Lorentzian metric satisfying the vacuum Einstein equations and exhibiting some Killing isometry. Let us restrict $\xi$ to define this Killing isometry, which is equivalent to the condition that the Lie derivative of the metric along $\xi$ vanishes:
\begin{equation}
(\mathcal{L}_\xi g)_{\mu\nu}=\xi^\rho g_{\mu \nu,\rho}+\xi^\rho,_\mu g_{\nu \rho}+\xi^\rho,_\nu g_{\mu \rho}=0. \label{killing}
\end{equation}
The twist potential
\begin{equation}
\omega_\mu=\sqrt{-\operatorname{det}(g)}\,\epsilon_{\mu\nu\sigma\rho}\xi^\nu\nabla^\sigma\xi^\rho, \label{twistpotential}
\end{equation}
and Killing vector norm
\begin{equation}
\lambda=-\xi^\mu\xi_\mu
\end{equation}
give the Ernst one-form
\begin{equation}
\sigma_\mu=\nabla_\mu \varsigma=\nabla_\mu\lambda-i\omega_\mu \label{ernstoneform}
\end{equation}
for some scalar $\varsigma$ (exactness is guaranteed by vanishing Ricci tensor, see \cite{Wald:1984rg} p.164).
Define a self-dual two form
\begin{equation}
\mathcal F_{\mu\nu}=(1+i*)\nabla_{[\mu} \xi_{\nu]} \, ,
\end{equation}
where $*$ is the Hodge dual operator. The \textit{spacetime} Ehlers group is defined for $(3+1)$-dimensional Lorentzian metrics by \eqref{ehler} for $W$ satisfying
\begin{subequations}\label{Wform}\begin{gather}
\nabla_{[\mu}W_{\nu]}=-2\gamma\Re[(\gamma \varsigma+i\delta)\mathcal F_{\mu\nu}]\label{Wforma}\\
\xi^\alpha W_\alpha+1 = (i\gamma\varsigma+\delta)(-i\gamma \bar{\varsigma}+\delta) \label{Wformb}
\end{gather}\end{subequations}
where a bar denotes complex conjugation, and $\gamma$ and $\delta$ are non-simultaneously vanishing real constants, which as a pair fix the gauge of $W$. The transformation defines an $SL(2,\mathbb{R})$ group action on the Ernst scalar by the M\"obius map
\begin{equation}
\varsigma\rightarrow\frac{\delta'\varsigma+i\gamma'}{i\gamma\varsigma+\delta},\qquad\text{where}\quad \gamma'\gamma+\delta'\delta=1. \label{sl2r}
\end{equation}

\section{Solution-generating transformations on the Navier-Stokes fluid} \label{sec:SGT in NS}

Consider those transformed metrics $h(\xi,W,g)$ which preserve the functional form $g$ (it clear that this is not in general the case). In the case of the Rindler metric dual to the incompressible Navier-Stokes fluid, we define the parameters of $g$ by the fluid velocity $v_i$, pressure $P$, and boundary position $r_c$ within the bulk. In the transformed metric $h$, we define the transformed parameters by $\w v_i$, $\w P$ and $\w r_c$, denoted by '$\sim$'.  On satisfying the vacuum Einstein equations on $\w\Sigma_c$, now at $r=\w r_c$ in the transformed geometry, the transformed metric will yield the incompressible Navier-Stokes equations in the transformed parameters
\begin{subequations}\begin{gather}
\partial_i\w v_i=0\\
\partial_\tau\w v_i+\partial_i\w P+\w v_k\partial_k\w v_i-\w r_c\partial^2\w v_i=0.
\end{gather}\label{Edual}\end{subequations}
Vitally, if $(v_i,P)$ satisfy the Navier-Stokes equations with viscosity $\nu=r_c$, then the transformed velocity and pressure $(\w v_i,\w P)$ represent a new set of solutions for viscosity $\nu=\w r_c$. That is, we look for a subset of the generalized Ehlers transformation acting on the fluid metric \eqref{g}, obeying some Killing isometry, which corresponds to solution-generating transformations of the velocity and pressure, and RG flow parametrised by $r_c$, of an incompressible Navier-Stokes fluid.

The Rindler metric is just one fluid metric supporting flat background geometries on the boundary. We therefore only wish instead to retain the common features of such metrics; the metric gauge $g_{\mu r}$, and the flat boundary metric of the form \eqref{flat metric}. The equation we wish to solve is thus
\begin{equation}
g_{\mu\nu}(r_c,v_i,P) \rightarrow h_{\mu \nu} (\xi,W,g)=\w g_{\mu\nu}=g_{\mu\nu}(\w r_c,\w v_i,\w P), \label{duality}
\end{equation}
where
\begin{subequations}
\begin{gather}
\w g_{\tau r}=1+\frac{\w v(x^a)^2+2\w P(x^a)}{2\w r_c},\qquad \w g_{ir}=-\frac{\w v_i(x^a)}{\w r_c},\qquad \w g_{rr}=0\\
\w g_{ab}|_{\w r_c}=\w\gamma_{ab},\qquad\text{where}\qquad	\w\gamma_{\tau\tau}=-\w r_c, \qquad \w\gamma_{ai}=\gamma_{ai}.
\end{gather}\label{dualmetric}
\end{subequations}

\subsection{Transforming the fluid}\label{sec:fluidtransformations}

We are provided in \eqref{dualmetric} with sufficient information to derive the possible fluid transformations via the form of the one-form $W$. Preserving the vanishing of the $\w g_{rr}=g_{rr}=0$ component of the metric we find, directly from \eqref{ehler}, the two possibilities
\begin{equation}
W_r=-2\alpha\xi_r\Omega^2/\lambda\qquad\text{where}\quad \alpha=0,1. \label{W_r}
\end{equation}
One may obtain an expression for $W_a$ by contraction of \eqref{ehler} with the boundary indices $(a,b,\ldots)$ of the Killing vector:
\begin{equation}
W_a=\frac{\Omega^2 \xi^r (g_{ar}+\xi_a\xi_r/\lambda)+\xi^b\w g_{ab}}{\lambda/\Omega^2+(1-2\alpha)\xi^r\xi_r}-\frac{\Omega^2\xi_a}{\lambda}.\label{Wa@wrc}
\end{equation}
(Note, here and in what follows $\xi_\mu= g_{\mu \nu} \xi^\nu$, ie. it is lowered with the metric $g_{\mu \nu}$ and never with $\tilde{g}_{\mu \nu}$). 
This expression is uniquely defined only at the dual boundary $\w\Sigma_c$, where we have defined the form of $\w g_{ab}$ and $W_a$ becomes independent of the dual fluid velocity and pressure. These expressions diverge for null Killing vectors, where $\lambda=0$. We cover this case shortly.

One can see that the parameters of the fluid is determined, to all orders in $\epsilon$, by $g_{ar}=g_{ar}|_{r_c}$. Consequently, the transformation in the fluid will be given by the transformation of these components. Evaluation at $\w r_c$ is necessary in order to circumvent the ambiguity in the dual metric, and also provides explicit fluid transformations. We begin with Killing vectors null at the dual boundary, $\lambda|_{\w r_c}=0$, where one finds from contraction of \eqref{ehler} with the Killing vector, which yields
\begin{equation}
\frac{\lambda}{\Omega^2}W_\mu=\xi^\nu(\w g_{\mu\nu}-g_{\mu\nu}),\label{ehler*xi}
\end{equation}
the following transformation,
\begin{equation}
\w g_{ar}=g_{ar}+\left[\frac{\xi^b(g_{ab}-\w\gamma_{ab})}{\xi^r}\right]_{\w r_c},\label{g_artransformation-null}
\end{equation}
accompanied by the preservation of a null Killing vector, $\xi^\mu\xi^\nu\w g_{\mu\nu}=0$.

For non-null Killing vectors we employ the relations
\begin{equation}
\xi^a(\w g_{ar}-(1-2\alpha)g_{ar})=0,\label{xidgar}
\end{equation}
derived by comparing \eqref{W_r} and \eqref{ehler*xi}, and
\begin{equation}
\lambda/\Omega^2=-\xi^\mu\xi^\nu\w g_{\mu\nu}, \label{ehler*xi*xi}
\end{equation}
found from contraction of \eqref{ehler} twice with the Killing vector. Inserting $W_r$ \eqref{W_r} and $W_a$ \eqref{Wa@wrc} into the Ehlers transformation \eqref{ehler} and employing \eqref{xidgar} and \eqref{ehler*xi*xi}, one finds
\begin{equation}
\w g_{ar}=\left[\frac{-\lambda g_{ar}+\xi_r((1-2\alpha)\xi^b\w\gamma_{ab}-\xi_a)}{\xi^c\xi^d\w \gamma_{cd}+(1-2\alpha)\xi^r\xi_r}\right]_{\w r_c}.\label{g_artransformation}
\end{equation}

\subsubsection{Energy scaling invariance from a bulk isometry}\label{sec:RG duality}

We begin with an example of a (null) Killing vector into the bulk, $\xi=\xi^r(x^\mu)\partial_r$. The Killing equation components $(\mathcal{L}_\xi g)_{ai}=0$ yield 
\begin{align}
\xi^r,{}_\tau &=\frac{1}{2}\xi^r\left(1+\frac{v^2+2P}{2r_c}+\mathcal{O}(\epsilon^4)\right),\\
\xi^r,{}_i &=-\xi^r\left(v_i\left(1-\frac{v^2+2P}{r_c}\right)+g^{(3)}_{i\tau,r}+\mathcal{O}(\epsilon^5)\right).
\end{align}
Integrability of these equations requires firstly
\begin{equation}
2v_j\partial_{[i}v_{j]}=-r_c\partial^2v_i+\mathcal{O}(\epsilon^5), \label{bulkconstraint}
\end{equation}
where we have used the Navier-Stokes equations to express the constraint in this form. Additionally, integrability requires vanishing vorticity to first order, which with incompressibility implies \eqref{bulkconstraint}. Transformation \eqref{g_artransformation-null} yields $\w g_{ar}=g_{ar}$, or
\begin{equation}
\w v_i = \frac{\w r_c}{r_c} v_i, \quad \w P=\frac{\w r_c}{r_c} P+\frac{\w r_c}{r_c}\left(1-\frac{\w r_c}{r_c}\right)\frac{v^2}{2},\label{bulk sgt}
\end{equation}
which is exact to all orders. It is trivial to show that the pair $(\w v_i,\w P)$ satisfy the incompressible Navier-Stokes equations (with viscosity $\w r_c$) if $(v_i,P)$ do so (with viscosity $r_c$) for velocities satisfying \eqref{bulkconstraint} alone - vanishing vorticity imposes unnecessary constraint and removes the dissipative term from the fluid equations.

It is interesting to consider the problems of existence, uniqueness and regularity of the Navier-Stokes in this case. The divergence of \eqref{bulkconstraint} yields a vanishing mean square vorticity which ensures the class of solutions $(v_i,P)$ generated by \eqref{bulk sgt} are regular. With respect to existence, the kinetic energy scales by a factor $\w r_c/r_c$ and thus is bounded if there exists any solution satisfying \eqref{bulkconstraint} where the energy is finite.

\subsubsection{The timelike Killing vector}

One might expect, in the presence of a timelike Killing vector $\xi=\partial_\tau$ (it is sufficient for this discussion to consider stationary solutions), a transformation of the form
\begin{equation}
\w v_i=-v_i,\qquad \w P=P,\qquad \w r_c=-r_c
\end{equation}
enacting time-reversal of the fluid but this is not the case. This is explained by noting that time-reversal is enacted by redefining the viscosity by $\nu=\pm r_c$ \cite{Bredberg:2011jq} rather than by changing $r_c$ itself. This is because sending $r_c\rightarrow -r_c$ brings the fluid outside the causal region of the spacetime.

\subsection{Fixed viscosity transformations}

We turn to fixed boundary (viscosity) transformations, where $\w r_c=r_c$. For Killing vectors null at the dual boundary then $\alpha=0$, and one recovers the identity. For non-null Killing vectors with $\alpha=1$, one finds
\begin{equation}
\w g_{ar}=g_{ar}-2\xi_r\left[\frac{\xi^b\gamma_{ab}-\xi^r g_{ar}}{\xi^c\xi^d\gamma_{cd}-\xi^r\xi_r}\right]_{r_c},\label{fixedrc}
\end{equation}
which defines a $\mathbb{Z}_2$ group.

\subsubsection{Spacelike Killing vectors} \label{sec:spacelike}

Consider a generic space-like Killing vector $\xi=\xi^k\partial_k$. Under \eqref{fixedrc}, the pressure is preserved, while the velocity transforms as
\begin{equation}
\w v_i=v_i-2\xi_i\frac{\sum_k\xi^kv_k}{\sum_j(\xi^j)^2}, \label{spaceliketransformation}
\end{equation}
which is a reflection in the hyperplane normal to the Killing vector and containing the point at which the velocity is defined.

\subsubsection*{Translational isometry}

Consider $\xi=c_k\partial_k$ where the constants $c_k$ are normalised to $\sum_k c_k^2=1$, and the corresponding isometries are $c_k\partial_k v_i=c_k\partial_k P=0$. The dual fields are
\begin{equation}
\w v_i=v_i-2c_i c_k v_k\qquad \w P=P.
\end{equation}
The incompressibility condition
\begin{align}
\partial_i \w v_i=\partial_i v_i-2c_i c_k \partial_i v_k=0,
\end{align}
and Navier-Stokes equations
\begin{equation}\begin{split}
\partial_\tau \w v_i+\partial_i\w P+\w v_k\partial_k\w v_i-r_c\partial^2\w v_i
=&(\delta_{ik}-2c_i c_k)(\partial_\tau v_k+\partial_k P+v^j\partial_j v_k-r_c\partial^2 v_k)\\
&+2c_ic_k\partial_k P-2c_jv_jc_k\partial_k(v_i-2c_ic_lv_l)=0,
\end{split}\end{equation}
are satisfied by the incompressible Navier-Stokes equations in the original fluid parameters along with the isometries. This is valid for fluids of arbitrary dimension.

\subsubsection*{Rotational isometry}

Consider a Killing vector $\xi=-x_2\partial_1+x_1\partial_2$ corresponding to a rotational isometry in a $d$-dimensional fluid.  In polar coordinates
\begin{equation}
x_1=\rho\cos\theta,\qquad x_2=\rho\sin\theta\qquad x_{k'}=x_{k'},
\end{equation}
where primed ($'$) indices run from $3$ to $(d-1)$, and the Killing vector becomes $\xi=\partial_\theta$, the isometries are
\begin{equation}
\partial_\theta v_1=-v_2 \qquad \partial_\theta v_2=v_1 \qquad \partial_\theta v_{k'}=0 \qquad \partial_\theta P=0. \label{angularisometries}
\end{equation}
Solutions satisfy
\begin{equation}
v_1=x_1\mu(\tau,x_{k'},\rho)-x_2\eta(\tau,x_{k'},\rho)\qquad v_2=x_2\mu(\tau,x_{k'},\rho)+x_1\eta(\tau,x_{k'},\rho)\label{angularvelocities}
\end{equation}
with dual velocities
\begin{equation}
\w v_1=x_1\mu(\tau,x_{k'},\rho)+x_2\eta(\tau,x_{k'},\rho)\qquad \w v_2=x_2\mu(\tau,x_{k'},\rho)-x_1\eta(\tau,x_{k'},\rho),
\end{equation}
that is, the transformation sends $\eta\rightarrow-\eta$ (equivalently $\theta\rightarrow-\theta$). The incompressible Navier-Stokes equations for the original fluid may be expressed as
\begin{subequations}
\begin{align}
0&=(2+\rho\partial_\rho)\mu+\partial_{k'}v_{k'}\\
0&=\begin{aligned}[t]\partial_\tau\mu+\rho^{-1}\partial_\rho P+\mu^2-\eta^2 +\mu\rho\partial_\rho\mu &+v_{k'}\partial_{k'}\mu\\
&-r_c(3\rho^{-1}\partial_\rho\mu+\partial_\rho^2\mu+\partial^{k'}\partial_{k'}\mu)\end{aligned}\\
0&=\partial_\tau\eta+2\mu\eta+\mu\rho\partial_\rho\eta+ v_{k'}\partial_{k'}\eta
-r_c(3\rho^{-1}\partial_\rho\eta+\partial_\rho^2\eta+\partial^{k'}\partial_{k'}\eta)\\
0&=\partial_\tau v_{k'}+\partial_{k'}P+\mu\rho\partial_\rho v_{k'}+v_{j'}\partial_{j'}v_{k'}-r_c\rho^{-1}\partial_\rho(\rho\partial_\rho v_{k'})
\end{align}
\end{subequations}
It is clear from the parity of these equations in $\eta$ that if there exists a fluid solution defined in terms of a pair $(\mu,\eta)$ by \eqref{angularvelocities}, then there also exists a solution parameterized by the pair $(\mu,-\eta)$. That is, the transformed fluid satisfies the incompressible Navier-Stokes equations. Again, this is valid for fluids of arbitrary dimension.

We provide an example with the three-dimensional fluid solution
\begin{subequations}
\begin{gather}
v_1=A\left(x_1-x_2e^{-2A(\tau-\tau_0)}\right)\qquad v_2=A\left(x_2+x_1e^{-2A(\tau-\tau_0)}\right)\\
v_3=B\exp\left(4A(\tau-\tau_0)+\frac{A\rho^2}{2r_c}\right)-e^{2A\tau}\int^\tau\ud\tau'\, q(\tau')e^{-2A\tau'}-2A x_3\\
P=\frac{1}{2}A^2\rho^2(e^{-4A(\tau-\tau_0)}-1)+q(\tau)x_3-2A^2 x_3^2
\end{gather}
\end{subequations}
which satisfies the isometries \eqref{angularisometries}. Here, $A$, $B$ and $\tau_0$ are arbitrary non-vanishing constants and $q(\tau)$ is an arbitrary function of time. The duality is equivalent to sending $\tau_0\rightarrow\tau_0+i\pi/2A$.

\subsection{Generalized versus spacetime Ehlers}\label{group for fixed rc}

In this section we discuss whether the fluid transformations of \S\ref{sec:spacelike}, derived directly from the generalized Ehlers transformation \eqref{ehler} on the basis of definition \eqref{dualmetric} of the dual metric components, lie within the spacetime Ehlers group. The spacetime Ehlers map is defined only in $(3+1)$ spacetime dimensions, so we assume this number of dimensions for all calculations in this section.

Consider the fixed boundary transformations \eqref{fixedrc}. For those examples calculated in \S\ref{sec:spacelike}, the Ernst scalar transforms as
\begin{equation}
\varsigma\rightarrow\bar\varsigma(1+\mathcal{O}(\epsilon^4)).
\end{equation}

In the case of the vacuum-to-vacuum spacetime Ehlers group, the Ernst scalar transforms according to the M\"obius map \eqref{sl2r}. If conjugation of the Ernst scalar is to belong to this map, we must have
\begin{equation}
\varsigma\rightarrow\bar\varsigma=\frac{\delta'\varsigma+i\gamma'}{i\gamma\varsigma+\delta},\qquad\text{where}\quad \gamma\gamma'+\delta'\delta=1,
\end{equation}
which is satisfied if and only if $\Omega^2=1$.

Explicit calculation for those examples in \S\ref{sec:spacelike} shows that $\Omega^2=1$ requires $(\gamma,\delta)=(0,\pm 1)$. The M\"obius map \eqref{sl2r} with these parameter values reduces to
\begin{equation}
\varsigma\rightarrow\varsigma\pm i\gamma'.
\end{equation}
That is, complex conjugation and the M\"obius map are equivalent only where they describe a constant complex shift. It follows from the definition of complex conjugation that this demands a vanishing twist potential \eqref{twistpotential}. For those fluid transformations of \S\ref{sec:spacelike}, the twist potential is non-vanishing, so complex conjugation of the Ernst scalar does not here correspond to the M\"obius map. Therefore these fluid transformations, generated by the generalized Ehlers group on the basis of the dual metric definition \eqref{dualmetric}, do not correspond to transformations within the spacetime Ehlers group.

\section{Discussion} \label{sec:discussion and outlook}

We have demonstrated how solution-generating transformations of the Einstein equations in the presence of a Killing vector may be applied to spacetime holographically dual to hydrodynamics. Our focus has been on the incompressible Navier-Stokes fluid dual to vacuum Rindler spacetime, where we have uncovered a selection of fluid transformations: a linear energy scaling symmetry for solutions with vanishing vorticity (this constraint may be relaxed to \eqref{bulkconstraint}), deriving from RG flow of the fluid hypersurface through the bulk, and a $\mathbb{Z}_2$ group of transformations for fixed viscosity (boundary) with explicit examples of reflection-like symmetry in translational and rotational fluid isometries. These transformations may not be remarkable from the perspective of hydrodynamics but it shows how part of the generalized Ehlers transformations can survive holography and give rise to transformations in the fluid dual. 

These fluid transformations, when applied to fluid metrics, will produce solution-generating transformations in the vacuum Einstein equations. However, the transformed metrics produced directly by our method are not necessarily vacuum. This apparent contradiction may be explained as follows. It is discussed in \cite{Niu:2011gu} how the electromagnetic field strength contribution to the Navier-Stokes equations \eqref{E} is determined by the projection of the electromagnetic field strength of the bulk spacetime along the unit normal to the hypersurface. If in the transformed spacetime this projected field strength vanishes, one will still recover the Navier-Stokes equations \eqref{Edual} (without forcing terms) in the dual parameters. In this way, it is not strictly necessary that the fluid metrics be vacuum to recover solution-generating transformations of the unforced incompressible Navier-Stokes equations.

One can then be inspired to try the Harrison transformation which is a solution-generating transformation in Einstein-Maxwell theory to give new transformations in magnetohydrodynamics. A holographic relation of the sort described here has been constructed for magnetohydrodynamics in \cite{magnetohydro, Niu:2011gu, Zou:2013ix}. The Harrison transformation in the bulk may then lead to nontrivial transformations in magnetohydrodynamics transforming between fluid velocity and magnetic potentials; this is the subject of current work. One can also study the dimensional dependence of these solution-generating transformations. In recent work \cite{Myers}, the difference in scaling for turbulence between three and four dimensions was studied holographically with a large difference in qualitative behaviour. One could also examine backgrounds associated with nontrivial chemical potential in the dual; for example rotating or charged black hole backgrounds.

Essentially, in this paper we wish to open up the use of gravitational solution generating symmetries in holography. It is encouraging that this did not immediately fail and one could preserve the fluid metric ansatz with some residual transformations surviving, yet it is intriguing that these transformations did not give anything particularly new. The results for magnetohydrodynamics may prove more significant.

\section*{Acknowledgements}

We would like to thank Cynthia Keeler and Andrew Strominger for their illuminating and insightful thoughts and comments on the fluid gravity correspondence and Shiraz Minwalla whose lectures on hydrodynamics at the Isaac Newton Institute programme on M-theory were an inspiration. This work is partially supported by STFC consolidated grant ST/J000469/1.


\begin{thebibliography}{99}



\bibitem{Damour}
	T. Damour,
	\emph{Quelques propri\'et\'es m\'ecaniques, \'electromagn\'etiques, thermodynamiques et quantiques des trous noirs},
	Th\`ese de Doctorat d'Etat, Universit\'e Pierre et Marie Curie, Paris VI, 1979.
\bibitem{Membrane Paradigm}
	K. S. Thorne, R. H. Price, D. A Macdonald,
	\emph{Black Holes: The Membrane Paradigm},
	Yale University Press, 1986
\bibitem{Baier:2007ix}
  R.~Baier, P.~Romatschke, D.~T.~Son, A.~O.~Starinets and M.~A.~Stephanov,
  \emph{Relativistic viscous hydrodynamics, conformal invariance, and holography},
  JHEP {\bf 0804} (2008) 100
  \href{http://arxiv.org/abs/0712.2451}{[arXiv:0712.2451 [hep-th]]}.
\bibitem{Kovtun:2004de}
  P.~Kovtun, D.~T.~Son and A.~O.~Starinets,
  \emph{Viscosity in strongly interacting quantum field theories from black hole physics},
  Phys.\ Rev.\ Lett.\  {\bf 94} (2005) 111601
  \href{http://arxiv.org/abs/hep-th/0405231}{[hep-th/0405231]}.
\bibitem{Policastro:2001yc}
  G.~Policastro, D.~T.~Son and A.~O.~Starinets,
  \emph{The Shear viscosity of strongly coupled N=4 supersymmetric Yang-Mills plasma},
  Phys.\ Rev.\ Lett.\  {\bf 87} (2001) 081601
  \href{http://arxiv.org/abs/hep-th/0104066}{[hep-th/0104066]}.


\bibitem{Bhattacharyya:2008jc}
  S.~Bhattacharyya, V.~EHubeny, S.~Minwalla and M.~Rangamani,
  \emph{Nonlinear Fluid Dynamics from Gravity},
  JHEP {\bf 0802} (2008) 045
  \href{http://arxiv.org/abs/0712.2456}{[arXiv:0712.2456 [hep-th]]}.
\bibitem{Bredberg:2010ky}
  I.~Bredberg, C.~Keeler, V.~Lysov and A.~Strominger,
  \emph{Wilsonian Approach to Fluid/Gravity Duality},
  JHEP {\bf 1103} (2011) 141
  \href{http://arxiv.org/abs/1006.1902v2}{[arXiv:1006.1902 [hep-th]]}.
\bibitem{Bredberg:2011jq}
  I.~Bredberg, C.~Keeler, V.~Lysov and A.~Strominger,
  \emph{From Navier-Stokes To Einstein},
  JHEP {\bf 1207} (2012) 146
  \href{http://arxiv.org/abs/1101.2451}{[arXiv:1101.2451 [hep-th]]}.
\bibitem{Compere:2011dx}
  G.~Compere, P.~McFadden, K.~Skenderis and M.~Taylor,
  \emph{The Holographic fluid dual to vacuum Einstein gravity},
  JHEP {\bf 1107} (2011) 050
  \href{http://arxiv.org/abs/1103.3022}{[arXiv:1103.3022 [hep-th]]}.
\bibitem{Bredberg:2011xw}
  I.~Bredberg and A.~Strominger,
  \emph{Black Holes as Incompressible Fluids on the Sphere},
  JHEP {\bf 1205} (2012) 043
  \href{http://arxiv.org/abs/1106.3084}{[arXiv:1106.3084 [hep-th]]}.
\bibitem{Cai:2011xv}
  R.~-G.~Cai, L.~Li and Y.~-L.~Zhang,
  \emph{Non-Relativistic Fluid Dual to Asymptotically AdS Gravity at Finite Cutoff Surface},
  JHEP {\bf 1107} (2011) 027
  \href{http://arxiv.org/abs/1104.3281}{[arXiv:1104.3281 [hep-th]]}.
\bibitem{Bhattacharyya:2008kq}
  S.~Bhattacharyya, S.~Minwalla and S.~R.~Wadia,
  \emph{The Incompressible Non-Relativistic Navier-Stokes Equation from Gravity},
  JHEP {\bf 0908} (2009) 059
  \href{http://arxiv.org/abs/0810.1545}{[arXiv:0810.1545 [hep-th]]}.
\bibitem{Compere:2012mt}
  G.~Compere, P.~McFadden, K.~Skenderis and M.~Taylor,
  \emph{The relativistic fluid dual to vacuum Einstein gravity},
  JHEP {\bf 1203} (2012) 076
  \href{http://arxiv.org/abs/1201.2678v2}{[arXiv:1201.2678 [hep-th]]}.
\bibitem{Eling:2012ni}
  C.~Eling, A.~Meyer and Y.~Oz,
  \emph{The Relativistic Rindler Hydrodynamics},
  JHEP {\bf 1205} (2012) 116
  \href{http://arxiv.org/abs/1201.2705v3}{[arXiv:1201.2705 [hep-th]]}.






\bibitem{Torre:1993jm}
  C.~G.~Torre and I.~M.~Anderson,
  \emph{Symmetries of the Einstein equations},
  Phys.\ Rev.\ Lett.\  {\bf 70} (1993) 3525
  \href{http://arxiv.org/abs/gr-qc/9302033}{[gr-qc/9302033]}.
\bibitem{Mars:2001gd}
  M.~Mars,
  \emph{Space-time Ehlers group: Transformation law for the Weyl tensor},
  Class.\ Quant.\ Grav.\  {\bf 18} (2001) 719
  \href{http://arxiv.org/abs/gr-qc/0101020}{[gr-qc/0101020]}.
\bibitem{Buchdahl:1959nk}
  H.~A.~Buchdahl,
  \emph{Reciprocal Static Metrics and Scalar Fields in the General Theory of Relativity},
  Phys.\ Rev.\  {\bf 115} (1959) 1325. \url{http://prola.aps.org/abstract/PR/v115/i5/p1325_1}.
\bibitem{Ehlers:1957zz}
  J.~Ehlers,
  \emph{Konstruktionen und Charakterisierung von Losungen der Einsteinschen Gravitationsfeldgleichungen}.
\bibitem{Geroch:1970nt}
  R.~P.~Geroch,
  \emph{A Method for generating solutions of Einstein's equations},
  J.\ Math.\ Phys.\  {\bf 12} (1971) 918. \href{http://dx.doi.org/10.1063/1.1665681}{doi:10.1063/1.1665681}.



\bibitem{Padmanabhan:2010rp}
  T.~Padmanabhan,
  \emph{Entropy density of spacetime and the Navier-Stokes fluid dynamics of null surfaces},
  Phys.\ Rev.\ D {\bf 83} (2011) 044048
  \href{http://arxiv.org/abs/1012.0119}{[arXiv:1012.0119 [gr-qc]]}.

\bibitem{Matsuo:2012pi}
  Y.~Matsuo, M.~Natsuume, M.~Ohta and T.~Okamura,
  \emph{The Incompressible Rindler fluid versus the Schwarzschild-AdS fluid},
  \href{http://arxiv.org/abs/1206.6924}{arXiv:1206.6924 [hep-th]}.




\bibitem{Chapman:2012my}
  S.~Chapman, Y.~Neiman and Y.~Oz,
  \emph{Fluid/Gravity Correspondence, Local Wald Entropy Current and Gravitational Anomaly},
  JHEP {\bf 1207} (2012) 128
  \href{http://arxiv.org/abs/1202.2469}{[arXiv:1202.2469 [hep-th]]}.



\bibitem{Huang:2011kj}
  T.~-Z.~Huang, Y.~Ling, W.~-J.~Pan, Y.~Tian and X.~-N.~Wu,
  \emph{Fluid/gravity duality with Petrov-like boundary condition in a spacetime with a cosmological constant},
  Phys.\ Rev.\ D {\bf 85}  (2012) 123531
  \href{http://arxiv.org/abs/1111.1576}{[arXiv:1111.1576 [hep-th]]}.

\bibitem{Rodrigues:2011gg}
  F.~G.~Rodrigues, W.~A.~Rodrigues, Jr and R.~da Rocha,
  \emph{The Maxwell and Navier-Stokes that Follow from Einstein Equation in a Spacetime Containing a Killing Vector Field},
  AIP Conf.\ Proc.\  {\bf 1483} (2012) 277
  \href{http://arxiv.org/abs/1109.5274}{[arXiv:1109.5274 [math-ph]]}.

\bibitem{Nakayama:2011bu}
  R.~Nakayama,
  \emph{The Holographic Fluid on the Sphere Dual to the Schwarzschild Black Hole},
  \href{http://arxiv.org/abs/arXiv:1109.1185}{arXiv:1109.1185 [hep-th]}.


\bibitem{Huang:2011he}
  T.~-Z.~Huang, Y.~Ling, W.~-J.~Pan, Y.~Tian and X.~-N.~Wu,
  \emph{From Petrov-Einstein to Navier-Stokes in Spatially Curved Spacetime},
  JHEP {\bf 1110} (2011) 079
  \href{http://arxiv.org/abs/1107.1464}{[arXiv:1107.1464 [gr-qc]]}.

\bibitem{Brattan:2011my}
  D.~Brattan, J.~Camps, R.~Loganayagam and M.~Rangamani,
  \emph{CFT dual of the AdS Dirichlet problem : Fluid/Gravity on cut-off surfaces},
  JHEP {\bf 1112} (2011) 090
  \href{http://arxiv.org/abs/1106.2577}{[arXiv:1106.2577 [hep-th]]}.


\bibitem{Kuperstein:2011fn}
  S.~Kuperstein and A.~Mukhopadhyay,
  \emph{The unconditional RG flow of the relativistic holographic fluid},
  JHEP {\bf 1111} (2011) 130
  \href{http://arxiv.org/abs/1105.4530}{[arXiv:1105.4530 [hep-th]]}.
















\bibitem{LL}
	L. D. Landau, E. M. Lifschitz,
	\emph{Fluid Mechanics, Course of Theoretical Physics}, Oxford: Pergamon Press, 1959.
\bibitem{Rangamani:2009xk}
  M.~Rangamani,
  \emph{Gravity and Hydrodynamics: Lectures on the fluid-gravity correspondence},
  Class.\ Quant.\ Grav.\  {\bf 26} (2009) 224003
  \href{http://arxiv.org/abs/0905.4352}{[arXiv:0905.4352 [hep-th]]}.


\bibitem{Symmetries of NS}
	V. N. Gusyatnikova, V. A. Yumaguzhin,
	\emph{Symmetries and conservation laws of navier-stokes
equations},
	Acta Applicandae Mathematicae \textbf{15} (January, 1989) 65-81.
\bibitem{Horvathy:2009kz}
  P.~A.~Horvathy and P.~-M.~Zhang,
  \emph{Non-relativistic conformal symmetries in fluid mechanics},
  Eur.\ Phys.\ J.\ C {\bf 65} (2010) 607
  \href{http://arxiv.org/abs/0906.3594}{[arXiv:0906.3594 [physics.flu-dyn]]}.










\bibitem{Woolley}
	M. L. Woolley
	\emph{On the role of the Killing tensor in the Einstein-Maxwell theory}
	Volume 33, Number 2 (1973), 135-144, \href{http://adsabs.harvard.edu/abs/1973CMaPh..33..135W}{doi: 10.1007/BF01645625}.
\bibitem{Harrison}
	B. K. Harrison,
	\emph{New Solutions of the Einstein-Maxwell Equations from Old}
	J. Math. Phys. 9, 1744 (1968); \href{http://dx.doi.org/10.1063/1.1664508}{doi: 10.1063/1.1664508}.
\bibitem{D Maison}
	D. Maison in 
	\emph{Duality and Hidden Symmetries in Gravitational Theories}, in
	\emph{Einstein's Field Equations and Their Physical Implications},
	B. G. Schmidt (ed.), Springer-Verlag, 2000.

\bibitem{Hoenselaers}
	H. Stephani, D. Kramer, M. MacCallum, C. Hoenselaers, E. Herlt,
	\emph{Exact Solutions of Einstein's Field Equations}, Second Edition. 732 p., Cambridge University Press, Cambridge, 2003. GBP80.00, ISBN 0521461367.
\bibitem{Stephani88}
	H. Stephani,
	\emph{Symmetries of Einstein's field equations with a perfect fluid source as examples of Li-Backl\"und symmetries},
	J. Math. Phys. 29, 1650 (1988); \href{http://dx.doi.org/10.1063/1.527913}{doi: 10.1063/1.527913}.
\bibitem{Garfinkle:1996ur}
  D.~Garfinkle, E.~N.~Glass and J.~P.~Krisch,
  \emph{Solution generating with perfect fluids},
  Gen.\ Rel.\ Grav.\  {\bf 29} (1997) 467
  \href{http://arxiv.org/abs/gr-qc/9611052v1}{[gr-qc/9611052]}.
\bibitem{Boonserm:2006vr}
  P.~Boonserm, M.~Visser and S.~Weinfurtner,
  \emph{Solution generating theorems for perfect fluid spheres},
  J.\ Phys.\ Conf.\ Ser.\  {\bf 68} (2007) 012055
  \href{http://arxiv.org/abs/gr-qc/0609088}{[gr-qc/0609088]}.
	
\bibitem{Wald:1984rg}
  R.~M.~Wald,
  \emph{General Relativity},
  Chicago, Usa: Univ. Pr. ( 1984) 491p.

\bibitem{magnetohydro}
  C.~-Y.~Zhang, Y.~Ling, C.~Niu, Y.~Tian and X.~-N.~Wu,
  \emph{Magnetohydrodynamics from gravity}
  Phys.\ Rev.\ D {\bf 86}  (2012) 084043
  \href{http://arxiv.org/abs/1204.0959}{arXiv:1204.0959 [hep-th]}.

\bibitem{Niu:2011gu}
  C.~Niu, Y.~Tian, X.~-N.~Wu and Y.~Ling,
  \emph{Incompressible Navier-Stokes Equation from Einstein-Maxwell and Gauss-Bonnet-Maxwell Theories},
  Phys.\ Lett.\ B {\bf 711} (2012) 411
  \href{http://arxiv.org/abs/1107.1430}{[arXiv:1107.1430 [hep-th]]}.

\bibitem{Zou:2013ix}
  D.~-C.~Zou, S.~-J.~Zhang and B.~Wang,
  \emph{The holographic charged fluid dual to third order Lovelock gravity},
 \href{http://arxiv.org/abs/1302.0904}{[arXiv:1302.0904 [hep-th].]}


\bibitem{Myers}
F.~Carrasco, L.~Lehner, R.~C.~Myers, O.~Reula and A.~Singh,
  \emph{Turbulent flows for relativistic conformal fluids in 2+1 dimensions}
  \href{http://arxiv.org/abs/1210.6702}{arXiv:1210.6702 [hep-th]}.

\end{thebibliography}
\end{document}